# Omni-tomography – Next-generation Biomedical Imaging

Ge Wang, Yue Wang, and Michael W. Vannier

December 21, 2012

Systems biomedicine represents one of the major efforts towards personalized and preventive medicine. Recent progresses are largely based on multi-platform and high-throughput –*omics* data from tissue samples and bioinformatics tools. There is a huge gap between these *in vitro* data and phenotypic *in vivo* features. We envision that futuristic research and its translation could greatly benefit from *in vivo* multi-platform, high-throughput and tomographic information about disease and pre-disease conditions. In-parallel to the development of novel multi-functional probes, multi-physics modeling, high-tech engineering and advanced image reconstruction present new opportunities to peek into living biological systems non-invasively without limitations in space and time. Building blocks are now either available or emerging for the birth of next-generation biomedical imaging – "*Omni-tomography*" (Wang, Zhang et al. 2012).

CT, MRI, PET, SPECT, ultrasound are all medical imaging modalities, each of which has a well-defined role. Over the past decade, we have seen an increasing popularity of multi-modality systems, such as PET-CT and PET-MR, gaining advantages by sequential or contemporaneous data acquisition (Cherry 2009; Patton, Townsend et al. 2009; van der Hoeven, Schalij et al. 2012). However, these paired modalities impose limitations that compromise our understanding of physiological processes relative to fine details and rapid changes driven by a beating heart. With omni-tomography, more or previously non-compatible imaging modalities can be fused together for a comprehensive study of local transient phenomena.

The first potential application is vulnerable plaque characterization, known as the "*holy grail*" of cardiology (Sanz and Fayad 2008; Vancraeynest, Pasquet et al. 2011; van der Hoeven, Schalij et al. 2012). Only in the omni-tomographic framework, CT and MRI can be now seamlessly merged for spatiotemporal registration, extendable to include other imaging modalities. CT provides structural definition and rapid snapshot. MRI reveals blood flow, soft tissue contrast, functional, cellular and molecular features. Neither CT nor MRI itself could cover all biomarkers of vulnerable plaques, including cap thickness, lipid-core size, stenosis, calcification, hemorrhage, elasticity, inflammation, endothelial status, oxidative stress, platelet aggregation, fibrin deposition, enzyme activity, microbial antigens, apoptosis and angiogenesis, and so on. It is exciting to have all of these features in high spatial, contrast, temporal resolution within a common coordinate system. Even if MRI of vulnerable plaques could have sufficient resolution and speed, retrospective image registration between CT and MRI is not a desirable alternative, because of registration errors due to non-repeatable contrast dynamics, organ motion and deformation, MRI-induced geometric distortion and signal nonlinearity, as well inconsistent contrast mechanisms between CT and MRI.

Intratumor heterogeneity evaluation is another example, found in many cancers after deep sequencing is applied to different parts of the same tumor (Costouros, Lorang et al. 2002; Segal, Sirlin et al. 2007; Cho, Ackerstaff et al. 2009; Albanese, Rodriguez et al. 2012; Shibata 2012). Solid tumors display remarkable phenotypic variability as related to angiogenesis, seed metastases, and therapeutic responses (O'Connor, Rose et al. 2011; Creixell, Schoof et al. 2012; Marusyk, Almendro et al. 2012). Advanced solid tumors often contain vascular compartments with distinct pharmacokinetics, comprising hypoxic regions and irregular vasculature (Jain 2005; Marusyk, Almendro et al. 2012). The more heterogeneous a tumor, the more likely therapy will fail the patient due to existence of drug-resistant phenotypes. Although a comprehensive characterization of intratumor heterogeneity at both genotypic and phenotypic levels cannot be achieved by any existing technology, omni-tomography opens a door for *in vivo* genotype-phenotype association and translational applications. It is desirable for ultrahigh resolution CT to depict a vascular tree, for diffusion tensor imaging to quantify local diffusion, and for contrast-enhanced MRI to capture circulatory dynamics. With omni-tomography, all these can be imaged in synchrony and analyzed for superior sensitivity and specificity; for example, to monitor tumor progression and treatment efficacy.

The enabling theory and technology for omni-tomography is "*interior tomography*" that has been developed over the past five years (Ye, Yu et al. 2007; Courdurier, Noo et al. 2008; Kudo, Courdurier et al. 2008; Yu and Wang 2009; Yang and et al. 2010; Katsevich, Katsevich et al. 2012). Traditional CT theory is for theoretically exact image reconstruction of a cross-section or a volume assuming projections are measured without any truncation. On the other hand, many important problems are local or at least often observed within a relatively small region of interest (ROI) such as the heart. The conventional wisdom is that an internal ROI cannot be exactly reconstructed only from truncated projection data associated with x-rays through the ROI (Natterer 1986). In 2007, it was proved that the interior problem can be exactly and stably solved if a sub-region in an ROI is known (Ye, Yu et al. 2007). The term "*interior tomography*" was then introduced to indicate the theoretical exactness of such an ROI reconstruction. Furthermore, it was analytically and experimentally showed that the interior problem permits a unique and stable solution if the

ROI is piecewise constant or polynomial (Yu, Cao et al. 2009; Yang and et al. 2010; Katsevich, Katsevich et al. 2012), which is a quite general image model (Wang and Yu 2010). Furthermore, interior tomography was recently elevated from its origin in CT to a general local tomographic image principle (Wang, Zhang et al. 2012). Its validity was already established for SPECT (Yu, Yang et al. 2009; Yang, Yu et al. 2012), MRI (Wang, Zhang et al. 2012), phase-contrast tomography (Cong, Yang et al. 2012; Yang, Cong et al. 2012), and others. As a result, relevant tomographic scanners can be made more compact, and integrated for comprehensive and simultaneous data acquisition from an ROI (Wang, Zhang et al. 2012).

Moving towards omni-tomography may start with "*simpler*" systems. As mentioned earlier, CT and MRI can be interiorized and combined for parallel imaging. This unification has the potential to reduce radiation dose when MRI-aided interior CT is implemented. On the other hand, CT-aided MRI can suppress motion artifacts. It is recognized that a rotating CT gantry can interfere with MRI. A solution is to use a stationary CT architecture in which multiple x-ray sources are around a patient along with the corresponding detector pieces (Wang, Yu et al. 2009). All the x-ray beams focus on an ROI, representing a few-view imaging setup. In the fused CT-MRI scanner, the synergy between CT and MRI data can be capitalized via compressive sensing (Gao, Yu et al. 2011; Wang, Zhang et al. 2012) and dictionary learning (Wang, Bresler et al. 2011). Since the CT gantry is not rotated, the electromagnetic shielding becomes much easier. While MRI is rather flexible in terms of imaging protocols, the latest spectral detector techniques promises to boost CT capabilities in terms of material decomposition and k-edge imaging (Walsh, Opie et al. 2011; Zainon, Ronaldson et al. 2012).

Omni-tomography offers biological, technical, physical, mathematical, and economic opportunities. Biologically, the "*all-in-one*" and "*all-at-once*" imaging power allows observation of well-registered spatiotemporal features *in vivo*. Physically, multi-physics modeling suggests new imaging modes for synergistic information (such as photoacoustic imaging which combines ultrasound resolution and optical contrast). Technically, a paradigm shift of system engineering is required to marry different types of imaging components. Economically, a "*one-stop-shop*" for diagnosis and intervention may be realized that could be often more cost-effective than a full-fledged imaging center with independent modalities. Omni-tomography does have limitations due to an ROI-oriented restriction, increased complexity and possible tradeoffs as more imaging contrast mechanisms are involved. Nevertheless, these are tractable with innovative technology and methods.

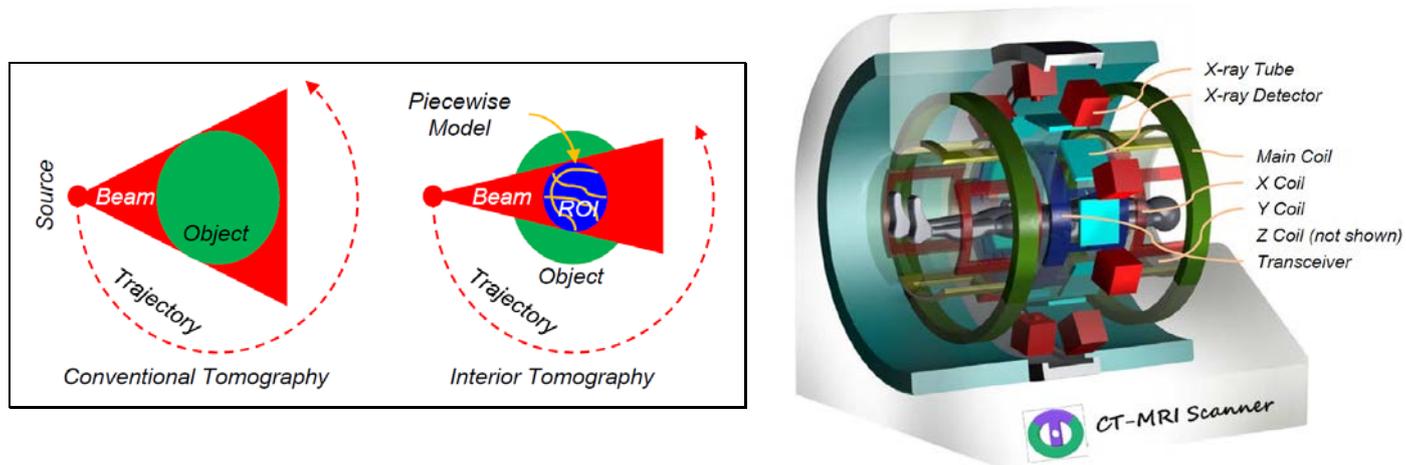

*Figure 1: From interior tomography to omni-tomography. Left:* While conventional tomography achieves exact global reconstruction of an object from a non-truncated scan, interior tomography targets exact region-of-interest (ROI) reconstruction from a truncated scan; *Right:* Omni-tomography is for grand fusion of multiple modalities, with an exemplary CT-MRI scanner design (rendered by Dr. Fenglin Liu with Biomedical Imaging Division at Virginia Tech) in which a pair of magnetic rings leaves space in the middle for CT, and potentially other modalities.